\documentclass[12pt]{article}

\usepackage{times}
\usepackage{epsf}
\usepackage[numbers,sort&compress]{natbib}
\usepackage{amsmath}
\usepackage{amsfonts}
\usepackage{amssymb}
\usepackage{bm}
\usepackage{bbm}
\usepackage{graphicx}
\usepackage{epsfig}
\usepackage{latexsym}
\usepackage{nicefrac}

\def\corresponds{{\lower.2ex\hbox{=}}{\rm\kern-.75em^\triangle}}
\def\succsim{\succ\kern-.9em_\sim\kern.3em}
\def\precsim{\prec\kern-1em_\sim\kern.3em}
\def\slantfrac#1#2{\kern1em^{#1}\kern-.3em/\kern-.1em_{#2}}
\def\lfrac#1#2{{}^{#1\!}\kern-.0em/_{#2}}

\def\buildrel#1\under#2{\mathrel{\mathop{\kern0pt #2}\limits_{#1}}}

\begin{document}
\bibliographystyle{myprsty}

\renewcommand{\thefootnote}{\fnsymbol{footnote}}

\vspace*{0.3cm}

\begin{center}
{\bf \Large Mass Measurements and the Bound--Electron $g$ Factor%
\footnote{Dedicated to Professor H.-J\"{u}rgen Kluge on 
the occasion of the 65$^{\rm th}$ birthday}}\\[2ex]
\normalsize U. D. Jentschura$^{a}$,
A. Czarnecki$^{b}$, K. Pachucki$^{c}$, and V. A. Yerokhin$^{d}$\\
\scriptsize
{\it \small
$^{a}$Max--Planck--Institut f\"ur Kernphysik,
  Saupfercheckweg 1, 69117 Heidelberg, Germany\\
$^{b}$Department of Physics, University of Alberta,
  Edmonton, AB, Canada T6G 2J1\\
$^{c}$Institute of Theoretical Physics, Warsaw University,
ul.~Ho\.{z}a 69, 00--681 Warsaw, Poland\\
$^{d}$Department of Physics, St.~Petersburg State University,
Oulianovskaya 1,\\ Petrodvorets, St.~Petersburg 198504, Russia}
\end{center}

\begin{abstract}
\noindent
The accurate determination of atomic masses and the
high-precision measurement of the bound-electron $g$ factor
are prerequisites for the determination of the electron mass,
which is one of the fundamental constants of nature.
In the 2002 CODATA adjustment [P. J. Mohr and B. N. Taylor,
Rev. Mod. Phys. {\bf 77}, 1 (2005)], the
values of the electron mass and the electron-proton mass ratio are
mainly based on $g$ factor measurements in combination with
atomic mass measurements. In this paper,
we briefly discuss the prospects for obtaining other fundamental
information from bound-electron $g$ factor measurements,
we present some details of a recent investigation of two-loop
binding corrections to the $g$ factor,
and we also investigate the radiative corrections
in the limit of highly excited Rydberg $S$ states
with a long lifetime, where the $g$ factor might be
explored using a double resonance experiment.\\[2ex]
{\bf PACS Nos.:} 31.30.Jv, 12.20.Ds, 11.10.St\\
{\bf Keywords:} Quantum Electrodynamics; Bound States; Atomic Physics.
\end{abstract}

%
%
\section{Introduction}

The central equation for the determination of the electron mass
$m_{\rm e}$ from $g$ factor measurements reads
\begin{equation}
\label{gbasic}
m_{\rm e} = \frac{\omega_{\rm c}}{\omega_{\rm L}} \,
\frac{g \,|e|}{2 q}\, m_{\rm ion}\,,
\end{equation}
where $\omega_{\rm c}$ is the cyclotron frequency of the ion, $\omega_{\rm
L}$ the Larmor spin precession frequency, $q$ the ion charge, and
$m_{\rm ion}$ its mass. The quantity $e = - |e|$ is the elementary
charge, and $g$ is the bound-electron $g$ factor.
In most practical applications, the ion is hydrogenlike,
and the frequency ratio $\omega_{\rm c}/\omega_{\rm L}$
can be determined very accurately in a Penning
trap~\cite{HaEtAl2000prl,VeEtAl2004}.

Equation~(\ref{gbasic}) may now be interpreted in different ways:
\begin{itemize}
\item The ratio $m_{\rm e}/m_{\rm ion}$ is immediately
accessible, provided we assume that quantum electrodynamic theory
holds for $g$. Provided the ratio $m_{\rm ion}/m_{\rm p}$
(with the proton mass $m_{\rm p}$) is also available to sufficient
accuracy, the electron to proton mass ratio $m_{\rm e}/m_{\rm p}$
can be determined by multiplication $m_{\rm e}/m_{\rm ion} \times
m_{\rm ion}/m_{\rm p}$. In the recent CODATA adjustment~\cite{MoTa2005},
the ratio $m_{\rm e}/m_{\rm p}$ has been determined
using two measurements involving ${}^{12}{\rm C}$.
\item Let us suppose that $m_{\rm ion}$ is known to sufficient
accuracy. Assuming that quantum electrodynamic theory
holds for $g$, we may then determine $m_{\rm e}$ from the
measurement~\cite{BeEtAl2002prl,PaJeYe2004,PaCzJeYe2005}.
\item The $g$ factor depends on the reduced mass of the
electron-ion two-particle system. An accurate measurement
of $g$ can therefore yield an independent verification
of the isotopic nuclear mass difference, provided that
the masses of the ions have been determined beforehand to sufficient
accuracy~\cite{BeEtAl2002}.
\item Direct access to the electron $g$ factor
in a weak external magnetic field depends on the
property of the nucleus having zero spin.
According to a relatively recent
proposal~\cite{WeEtAl2000,MoEtAl2005},
the measurement of a $g$ factor for a nucleus with non-zero
spin can be used to infer the nuclear $g$ factor,
provided the purely electronic part of the $g$ factor
is known to sufficient accuracy from other measurements.
\item There is also a proposal for measuring $g$ factors
in lithiumlike systems, and theoretical work in this direction has
been undertaken~\cite{ShEtAl2002}. Provided the contribution due to
electron-electron correlation can be tackled to sufficient accuracy,
a measurement of the $g$ factor in lithiumlike systems could give
access to the nuclear-size effect, which in turn can be used as an
additional input for other determinations of fundamental constants.
\item Finally, provided the mass $m_{\rm ion}$ of a high-$Z$ ion
is known to sufficient accuracy
and $m_{\rm e}$ is taken from $g$ factor measurements at lower
nuclear charge number, the high-$Z$ experimental result for $g$ may be
compared to a theoretical prediction, yielding a test of
quantum electrodynamics for a bound particle subject to an
external magnetic field and a strong Coulomb field
(see, e.g., Sec.~2.2 of~\cite{SPARC}).
Alternatively, one may invert the relation $g = g(\alpha)$
to solve for the fine-structure constant (important precondition:
knowledge of nuclear size effect)~\cite{WeEtAl2000,Ka2001proc}.
The feasibility of the
latter endeavour in various ranges of nuclear charge numbers
will be discussed in the current article.
\end{itemize}
These examples illustrate the rich physics implied by
$g$ factor measurements in combination with the determination
of atomic masses via Penning traps. Indeed, the $g$ factor
is a tremendous source of information regarding fundamental
constants, fundamental interactions and nuclear properties.

This paper is organized as follows.
In Sec.~\ref{atmmass}, we briefly discuss the importance and
the status of
atomic mass measurements for further advances. In
Sec.~\ref{theory}, we describe a few details of two recent
investigations~\cite{PaJeYe2004,PaCzJeYe2005} regarding
one- and two-loop binding corrections to the $g$ factor, and in
Sec.~\ref{asymp}, we discuss the asymptotics of the
corrections for high quantum numbers, with a partially
surprising result, before dwelling on connections of the
$g$ factor to nuclear effects and the fine-structure constant
in Sec.~\ref{nucl}. Conclusions are drawn in Sec.~\ref{conclu}.
An Appendix is devoted to the
current status of the free-electron anomaly.

\section{Atomic Mass Measurements -- Present and Future}
\label{atmmass}

A review of the current status of atomic mass measurements
can be in found in Ref.~\cite{AuWaTh2003}.
Experimental details regarding modern atomic mass
measurements, with a special emphasis on hydrogenlike ions,
can be found in Refs.~\cite{BeEtAl2003epjd,KlEtAl2003}.
Regarding the current status of mass measurements,
one may point out that some of the masses of S, Kr and Xe ions
have recently been determined with an accuracy
of better than 1 part in $10^{10}$ (Ref.~\cite{ShReMy2005}).
For molecular ions, the accuracy has recently been pushed
below $10^{-11}$~\cite{RaThPr2004}.

Recent measurements for the hydrogenlike ions
${}^{24}{\rm Mg}^{11+}$ and ${}^{26}{\rm Mg}^{11+}$
(Ref.~\cite{BeEtAl2003epjd}) and ${}^{40}{\rm Ca}^{19+}$
(Ref.~\cite{ScPriv2005}), as well as for the
lithiumlike ion ${}^{40}{\rm Ca}^{17+}$
(Ref.~\cite{ScPriv2005}) have reached an accuracy
of about $5 \times 10^{-10}$. These experiments pave the way for
accurate determinations of fundamental
constants using $g$ factor measurements in these systems.
At the University of Mainz~\cite{BlPriv2005} (MATS collaboration)
and at the University of Stockholm~\cite{ScPriv2005}
(SMILE-TRAP), there are plans to significantly extend and
enhance atomic mass measurements (including many more isotopes
and nuclei) over the next few years, with accuracies below 1 part in
$10^{11}$ or even $10^{12}$. Eventually,
one may even hope to determine the nuclear size effect
of a specific ion by ``weighing'' the Lamb shift of the ground state.
In the same context, one may point out that the masses of
different charge states of ions are determined vice versa
by adding and subtracting binding energies. This implies, e.g.,
that the mass of ${}^{12}{\rm C}^{5+}$ in terms of the
mass of neutral carbon,
$m({}^{12}{\rm C}) = 12\,{\rm u}$, is given by
\begin{equation}
m({}^{12}{\rm C}^{5+}) = m({}^{12}{\rm C}) - 5 \, m_{\rm e}
  + c^{-2} \, E_{\rm B} \,,
\end{equation}
where $E_{\rm B} = 579.835(1) \times 10^{-9} \, {\rm u} c^2$ is the
cumulative binding energy for all 5 electrons~\cite{MoTa2000}.
This relation has proven useful in the determination of the
electron mass~\cite{BeEtAl2002}.

In order to make a comparison to the accuracy of the free-electron
determination of $\alpha$, it is perhaps useful to remember that
in the seminal work~\cite{DyScDe1987},
the free-electron and positron anomaly has been
determined to an accuracy $4 \times 10^{-9}$.
This translates into a level of accuracy
of about $4 \times 10^{-12}$ for the $g$ factor itself.
The accuracy of the current value of $\alpha$ is
$4 \times 10^{-9}$~\cite{MoTa2005}.

\section{Calculation of the Bound--Electron $g$ Factor}
\label{theory}

The bound-electron $g$ factor measures the
energy change of a bound electron (hydrogenlike ion,
spinless nucleus) under a quantal change in the
projection of the total angular momentum with respect to an
axis defined by a (weak) external magnetic field.
In this sense, the $g$ factor of a bound electron should rather
be termed the $g_J$ factor (according to the Land\'{e} formulation).

However, for $S$ states, the total angular momentum number is equal
to the spin quantum number, and therefore it has been common
terminology not to distinguish the notation for $g$ and $g_J$.

For a general hydrogenic state, the
Dirac-theory $g$ factor, denoted
$g_{\rm D}$, reads (see~\cite{MoEtAl2005} and references therein)
\begin{equation}
\label{gD}
g_{\rm D} = \frac{\kappa}{j(j+1)} \,
\left( \kappa\,\frac{E_{n\kappa}}{m_{\rm e}} - \frac12\right)\,.
\end{equation}
Here, $E_{nj}$ is the Dirac energy, and the quantum numbers
$n$, $j$ and $\kappa$ have their usual meaning.
Throughout this article, we use natural units with 
$\hbar = c = \epsilon_0 = 1$.

For $S$, $P$ and $D$ states, Eq.~(\ref{gD}) leads to the
following expressions (we here expand the bound-state energy
in powers of $Z\alpha$),
\begin{subequations}
\label{varg}
\begin{eqnarray}
\label{gnS12}
g_{\rm D}(nS_{1/2}) &=&  2 - \frac{2\,(Z\alpha)^2}{3\,n^2} -
\frac{(Z\alpha)^4}{n^3}\,
\left( \frac23 - \frac{1}{2\,n} \right) \,,\\
\label{gnP12}
g_{\rm D}(nP_{1/2}) &=&  \frac23 - \frac{2\,(Z\alpha)^2}{3\,n^2} -
\frac{(Z\alpha)^4}{n^3}\,
\left( \frac23 - \frac{1}{2\,n} \right) \,,\\
\label{gnP32}
g_{\rm D}(nP_{3/2}) &=& \frac43 - \frac{8\,(Z\alpha)^2}{15\,n^2} -
\frac{(Z\alpha)^4}{n^3}\,
\left( \frac{4}{15} - \frac{2}{5\,n} \right) \,,\\
\label{gnD32}
g_{\rm D}(nD_{3/2}) &=& \frac45 - \frac{8\,(Z\alpha)^2}{15\,n^2} -
\frac{(Z\alpha)^4}{n^3}\,
\left( \frac{4}{15} - \frac{2}{5\,n} \right) \,,\\
\label{gnD52}
g_{\rm D}(nD_{5/2}) &=& \frac65 - \frac{18\,(Z\alpha)^2}{35\,n^2} -
\frac{(Z\alpha)^4}{n^3}\,
\left( \frac{6}{35} - \frac{27}{70\,n} \right) \,.
\end{eqnarray}
\end{subequations}
The above formulas illustrate the in principle well-known
fact that the bound-electron
$g$ factor would be different from the free-electron
Dirac value $g=2$, even for $S$ states and even in the
absence of quantum electrodynamic loop corrections.

We now briefly summarize the
results of recent investigations~\cite{PaJeYe2004,PaCzJeYe2005}
of the bound-electron $g$ factor, which is based on
nonrelativistic quantum electrodynamics
(NRQED). The central result of this investigation is the
following semi-analytic expansion
in powers of $Z\alpha$ and $\ln(Z\alpha)$
for the bound-electron $g$ factor ($n{S}$ state)
in the non-recoil and pointlike-nucleus limit
(for recoil effects see e.g.~Ref.~\cite{Sh2001}):
\begin{align}
\label{gbound}
g(n{S}) & =
\underbrace{2 - \frac{2\, (Z\alpha)^2}{3\,n^2} +
\frac{(Z\alpha)^4}{n^3} \, \left( \frac{1}{2 n} - \frac23 \right) +
{\cal O}{(Z\alpha)^6}}_{\mbox{Breit (1928),
Dirac theory}}
\nonumber\\[4ex]
&
+ \underbrace{{\frac{\alpha}{\pi}}\,
\left\{ {2 \times \frac{1}{2}}\, {\bigg(}
1 + \frac{(Z\alpha)^2}{6 n^2} {\bigg)}
+ \frac{(Z\alpha)^4}{n^3} {\bigg\{}
a_{41}\, {\ln}[(Z\alpha)^{-2}]  +
a_{40} {\bigg\}}
+ {\cal O}(Z\alpha)^5 \right\}}_{\mbox{one-loop correction}}
\nonumber\\[4ex]
&
+ \underbrace{{ \left(\frac{\alpha}{\pi}\right)^2}\,
\left\{ {-0.656958} \,  {\bigg(} 1 +
\frac{(Z\alpha)^2}{6 n^2} {\bigg)}  +
\frac{(Z\alpha)^4}{n^3} {\bigg\{}
b_{41}\,
{\ln}[(Z\alpha)^{-2}] +
b_{40} {\bigg\}}
+ {\cal O}(Z\alpha)^5 \right\}}_{\mbox{two-loop correction}}
\nonumber\\[4ex]
& + {\cal O}(\alpha^3)\,.
\end{align}
This expansion is valid through the order
of two loops (terms of order $\alpha^3$ are
neglected). The notation is in part inspired by the usual conventions
for Lamb-shift coefficients~\cite{SaYe1990}:
the (lower case) $a$ terms denote the
one-loop effects, with $a_{kj}$ denoting the
coefficient of a term proportional to
$\alpha\,(Z\alpha)^k\,{\ln}^j[(Z\alpha)^{-2}]$.
The $b$ terms denote the two-loop corrections,
with $b_{kj}$ multiplying
a term proportional to
$\alpha^2\,(Z\alpha)^k\,{\ln}^j[(Z\alpha)^{-2}]$.
In~\cite{PaJeYe2004,PaCzJeYe2005}, complete results are derived for the
coefficients $a_{41}$, $a_{40}$, $b_{41}$ and $b_{40}$,
valid for arbitrary excited $S$ states in hydrogenlike systems.

\begin{figure}[htb]
\begin{center}
\begin{minipage}{14cm}
\begin{center}
\includegraphics[width=0.7\linewidth]{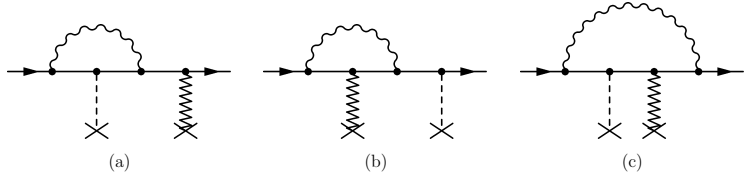}
\caption{\label{fig1} One--loop, two-vertex scattering
diagrams that correspond to the one-loop part of the effective operators
Eq.~(B16) and~(B17) of Ref.~\cite{PaCzJeYe2005}. The zigzag line
denotes the interaction with the external field, whereas the
dashed lines denote the Coulomb photons. The two-loop
part of these effective operators is generated by diagrams with
one more virtual photon, with electron-photon vertices
to be inserted at all topologically distinguishable
positions in the electron lines of diagrams (a), (b) and (c).}
\end{center}
\end{minipage}
\end{center}
\end{figure}

In Eq.~(\ref{gbound}), the term underlined by ``Breit (1928),
Dirac theory'' corresponds to the prediction of relativistic
atomic theory, including the relativistic corrections
to the wave function~\cite{Br1928}. By contrast, the term
$\frac{\alpha}{\pi} \, (2 \times \frac{1}{2})$ in the
expression underlined by ``one-loop correction''
gives just the leading (Schwinger) correction to the
anomalous magnetic moment of a free electron.
This latter effect is modified here by additional
binding corrections to the one-loop correction,
which give rise e.g.~to terms of order $\alpha\,(Z\alpha)^2$
and higher (in $Z\alpha$). Perhaps, it is also worth
clarifying that the term $-0.656958$ is just twice the
two-loop contribution to the anomalous magnetic moment of a free
electron, which is usually quoted as $(\frac{\alpha}{\pi})^2 \, (-0.328479)$
in the literature.

Explicit results for the coefficients in (\ref{gbound}),
restricted to the one-loop self-energy, read~\cite{PaJeYe2004}
\begin{subequations}
\label{a4}
\begin{align}
\label{a41}
a_{41}(nS) &= \frac{32}{9}\,, \\[1ex]
\label{a40}
a_{40}(nS) &= \frac{73}{54}
- \frac{5}{24 n}
- \frac{8}{9} \, \ln k_0(n{S})
- \frac{8}{3} \, \ln k_3(n{S})\,.
\end{align}
\end{subequations}
Here, $\ln k_0(nS)$ is the Bethe logarithm for an $nS$ state,
and $\ln k_3(nS)$ is a generalization of the Bethe logarithm
to a perturbative potential of the form $1/r^3$
(see also Table~\ref{lnk3table} below).
Vacuum polarization adds a further $n$-independent
contribution of $(-16/15)$ to $a_{40}$~\cite{Ka2000}.
Higher-order binding corrections to the one-loop self-energy
contribution to the $g$ factor
have been considered, e.g., in~\cite{YeInSh2002},
and for the vacuum-polarization contribution,
one may consult, e.g., Ref.~\cite{BeEtAl2000pra}.

The results for the two-loop coefficients read
\begin{subequations}
\label{b4}
\begin{eqnarray}
\label{b41}
b_{41}(nS) &=& \frac{28}{9}\,, \\[1ex]
\label{b40}
b_{40}(nS) &=& \frac{258917}{19440} - \frac{4}{9}\,\ln k_0
   - \frac{8}{3}\,\ln k_3(nS) + \frac{113}{810}\,{\pi }^2
   -  \frac{379}{90}\,{\pi }^2\,\ln 2 + \frac{379}{60}\,\zeta(3)
\nonumber \\
&& + \frac{1}{n}\left(
   -\frac{985}{1728}   - \frac{5}{144}\,{\pi }^2 +
      \frac{5}{24}\,{\pi }^2\,\ln 2 - \frac{5}{16}\,\zeta(3)\right)\,.
\end{eqnarray}
\end{subequations}
Our result for $b_{40}$ includes contributions from all
two-loop effects (see Fig.~21 of~\cite{Be2000} for the diagrams)
up to the order $\alpha^2\,(Z\alpha)^4$.
The logarithmic term $b_{41}$ is, however,
exclusively related to the two-loop self-energy
diagrams. An essential contribution to the one- and two-loop
effects is given by two--Coulomb--vertex scattering
amplitudes (see also Fig.~\ref{fig1}).

\begin{figure}[htb]
\begin{center}
\begin{minipage}{14cm}
\begin{center}
\includegraphics[width=0.7\linewidth]{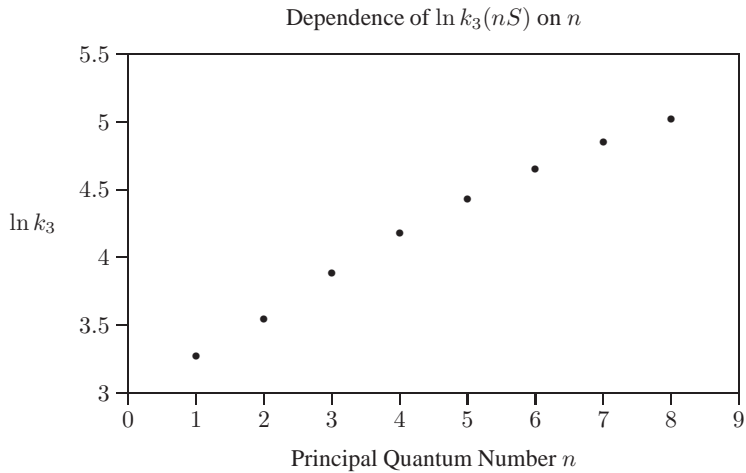}
\caption{\label{fig2} A plot of the generalized Bethe logarithms
$\ln k_3(nS)$ as a function of the principal quantum number $n$
illustrates the monotonic increase with $n$.
For the numerical values, see Table~\ref{lnk3table}.}
\end{center}
\end{minipage}
\end{center}
\end{figure}

%
%
\section{Asymptotics for High Quantum Numbers}
\label{asymp}

It is interesting to study the limit of the coefficients
$a_{40}$ and $b_{40}$ in the limit of highly excited states,
$n \to \infty$. For the Bethe logarithm $\ln k_0$, such a
study has recently been completed
(see Refs.~\cite{Po1981,JeMo2005bethe}).
The asymptotics of the generalized Bethe logarithm $\ln k_3$
have not yet been determined. We here supplement
the numerical result for $8S$. In Eq.~(72)
of Ref.~\cite{PaCzJeYe2005}, results have been communicated
for $S$ states with $n \leq 7$.

\begin{table}[htb]
\begin{center}
\begin{minipage}{10cm}
\begin{center}
\caption{\label{lnk3table}
A table of generalized Bethe logarithms $\ln k_3(nS)$ for
excited $S$ states. This quantity enters into
Eqs.~(\ref{a40}) and~(\ref{b40})
and characterize the one-loop binding correction to the
$g$ factor of order $\alpha\,(Z\alpha)^4$ and the two-loop
correction of order $\alpha^2\,(Z\alpha)^4$. All decimals
shown are significant.}
\begin{tabular}{cr}
\hline
\hline
$n$ & $\ln k_3(nS)$ \\
\hline
1 & 3.272~806~545\\
2 & 3.546~018~666\\
3 & 3.881~960~979\\
4 & 4.178~190~961\\
5 & 4.433~243~558\\
6 & 4.654~608~237\\
7 & 4.849~173~615\\
8 & 5.022~275~220\\
\hline
\hline
\end{tabular}
\end{center}
\end{minipage}
\end{center}
\end{table}

\begin{figure}[htb]
\begin{center}
\begin{minipage}{14cm}
\begin{center}
\includegraphics[width=0.5\linewidth]{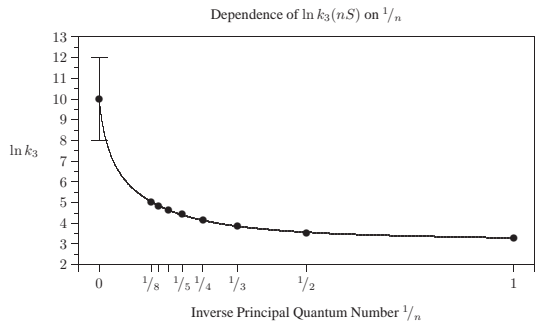}
\caption{\label{fig3} A plot of the generalized Bethe logarithms
$\ln k_3(nS)$ as a function $1/n$ instead of $n$ indicates
consistency with an asymptotic limit $\lim_{n \to \infty} \ln k_3(nS)
= 10 \pm 2$.}
\end{center}
\end{minipage}
\end{center}
\end{figure}

The result for $n=8$ confirms the trend of a monotonic increase
of $\ln k_3$ with $n$ (see Fig.~\ref{fig2}).
On the other hand, based on the general experience
regarding the structure of radiative corrections in the limit $n \to \infty$,
we would expect a constant limit of $\ln k_3(nS)$ for $n \to \infty$.
Using an extrapolation scheme similar to the one employed
in~\cite{LBEtAl2003}, we conjecture the following limit
(see Fig.~\ref{fig3}),
\begin{equation}
\lim_{n \to \infty} \ln k_3 (nS) = 10 \pm 2\,,
\end{equation}
It would be very interesting to verify this limit by an explicit
calculation, e.g., using the techniques outlined in Ref.~\cite{Po1981}.

Highly excited Rydberg states are characterized by a long lifetime.
In a Penning trap, however, the confining electric fields would tend to quench
transitions to lower-lying levels. One might attempt a measurement
of a $g$ factor of a Rydberg state
via a double-resonance approach, with one laser driving the spin flip
(Larmor precession frequency)
and another being tuned to a transition between Rydberg
states~\cite{QuPriv2005}.

%
%
\section{Bound--Electron $g$ Factor, Nuclear Effects and
the Fine--Structure Constant}
\label{nucl}

\begin{figure}[htb]
\begin{center}
\begin{minipage}{14cm}
\begin{center}
\includegraphics[width=1.0\linewidth]{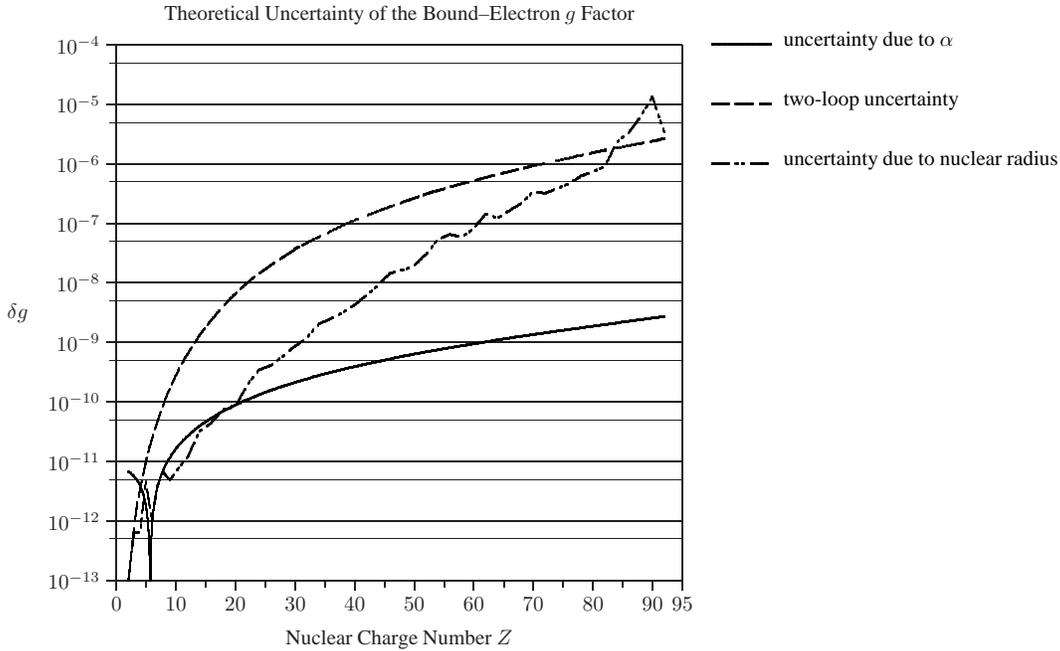}
\caption{\label{fig4} Various sources of theoretical uncertainty
for the bound-electron $g$ factor, over the entire $Z$ range
from hydrogen to uranium.}
\end{center}
\end{minipage}
\end{center}
\end{figure}

\begin{figure}[htb]
\begin{center}
\begin{minipage}{14cm}
\begin{center}
\includegraphics[width=1.0\linewidth]{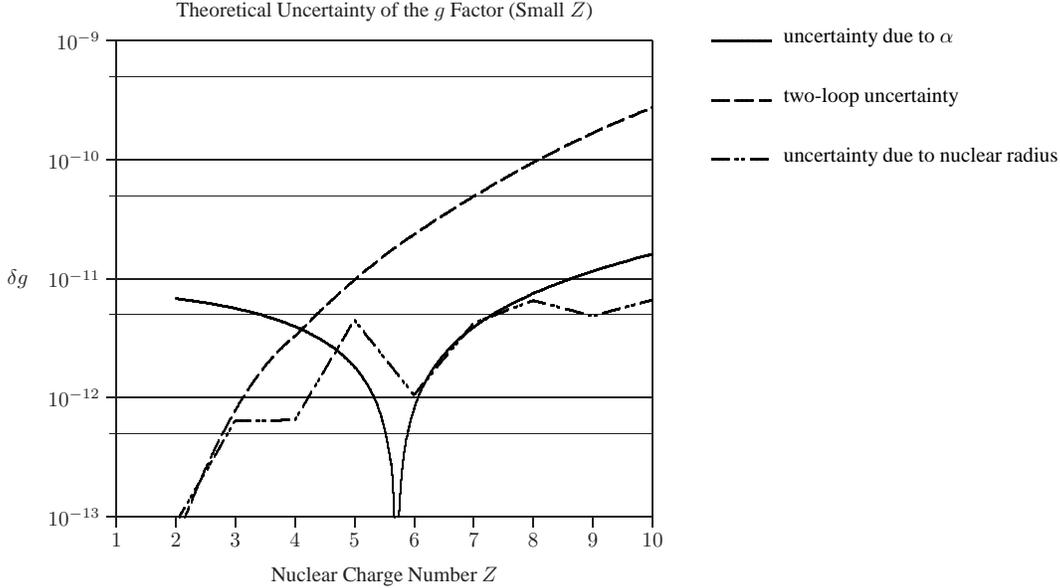}
\caption{\label{fig5} A close-up of Fig.~\ref{fig4} in the range
of small quantum numbers $n$ illustrates that an alternative
determination of the fine-structure constant, based on
the current theoretical status, would be possible
for ionized helium (${}^4 {\rm He}^+$) and
beryllium (${}^{10} {\rm Be}^+$).
The ${}^{6,7} {\rm Li}$ nuclei are not spinless.}
\end{center}
\end{minipage}
\end{center}
\end{figure}

In Figs.~\ref{fig4} and~\ref{fig5}, we indicate three
primary sources of the theoretical uncertainty of the bound-electron
$g$ factor across the entire range of nuclear charge numbers
(these are the fine-structure constant, higher-order unknown two-loop
effects and the nuclear radius). 
For a determination of the
fine-structure constant using the bound-electron $g$ factor,
the experimental accuracy would have to be improved
to a value below the corresponding uncertainty curve
in Figs.~\ref{fig4} and~\ref{fig5}. Such a determination
would constitute a very important and attractive
additional pathway, using bound-state quantum electrodynamics,
as an alternative to the ``usual'' determination based
on the free-electron $g$ factor.

Before we dwell further on the fine-structure constant,
we briefly discuss the
nuclear polarizability correction to the $g$ factor
(see Ref.~\cite{NePlSo2002} and Appendix~\ref{diamagpol})
which represents an additional obstacle in the determination
of the fine-structure constant from $g$ factor measurements.
One might hope that it can be accurately understood one day
in terms of nuclear models. In Appendix~\ref{diamagpol},
we present an additional nuclear effect (a magnetic 
susceptibility correction) which may also have to be taken 
into account in an accurate description of the 
nuclear contributions to the bound-electron $g$ factor,
especially in the range of medium nuclear charge numbers.  

The further shift/uncertainty of the $g$ factor, caused by the nuclear
finite-size effect (nuclear volume effect),
is typically smaller than the uncertainty
of the theoretical prediction for the $g$ factor due to higher-order
quantum electrodynamic two-loop binding corrections
(see Figs.~\ref{fig4} and~\ref{fig5}).
In evaluating the uncertainty due to the nuclear
radius, we have used the most recent values for
the root-mean-square (rms) nuclear radii~\cite{An2004}.

In order to investigate the sensitivity of the
bound-electron $g$ factor to the fine-structure constant,
we approximate the $g$ factor by the first two terms in
the $Z\alpha$-expansion of the Dirac theory and the
one-loop correction, and obtain
\begin{equation}
\label{dgdalpha}
\delta g \approx \left\{ -\frac23\,Z^2\,\alpha\,
\left[ 2 + (Z\alpha)^2 \right] +
\frac{1}{\pi}\, \left[1 + \frac12\,(Z\alpha)^2 \right] \right\}
\, \delta\alpha\,.
\end{equation}
For a determination of $\alpha$, it is desirable, in principle, to
tune the parameters so that the modulus $|\delta g|$ for
given $\delta \alpha$ becomes as large as possible.

For nuclear charge numbers in the (fictitious) range $5 \leq Z \leq 6$,
the sensitivity of $g$ on $\alpha$ suffers from a
cancellation of the one-loop against the Dirac binding corrections
(see also Figs.~\ref{fig4} and~\ref{fig5}), and we have
\begin{equation}
\frac{\delta g}{\delta\alpha} \approx 0 \qquad \mbox{for} \qquad
Z \approx 5.7\,.
\end{equation}
It would be rather difficult to determine $\alpha$ via a measurement
of the $g$ factor in the indicated range of nuclear charge numbers.

For large $Z$, one may find a crude
approximation to Eq.~(\ref{dgdalpha}) by the relation
\begin{equation}
\frac{|\delta g|}{\delta\alpha} \approx \frac43\,Z^2\,\alpha
\qquad \Rightarrow \qquad
\frac{|\delta g|}{g} \approx \frac23\,(Z\,\alpha)^2\,
\frac{\delta\alpha}{\alpha} \,.
\end{equation}
The enhancement of the theoretical uncertainty of $g$ with $Z$
is manifest in Fig.~\ref{fig4}.
In principle, one might assume that a measurement at high $Z$
could be more favourable for a determination of
$\alpha$ than a corresponding experiment in a low-$Z$ system.
However, as shown in Fig.~\ref{fig4}, the nuclear
structure alone currently entails an uncertainty of $g$ that is larger
than the uncertainty due to the fine-structure constant, for
large $Z$. Also, the uncertainty due to higher-order
unknown two-loop binding corrections currently represents
an obstacle for an alternative determination of $\alpha$
from a $g$ factor measurement at high $Z$.

Reversing the argument, one may point out that,
provided the two-loop uncertainty of the theoretical
prediction for large $Z$ can be
reduced substantially, one may infer the nuclear radius from the
measurement of the $g$ factor. Again, going one step further
and assuming that the nuclear radius is accurately known from
other measurements, e.g., Lamb shift experiments or
$g$ factor measurements in lithiumlike systems, one may
eventually hope to infer the fine-structure constant from a high-$Z$
measurement. This endeavour can thus be interpreted as a
rather difficult combined effort of theory and experiment,
with results not to be expected in the immediate future, but
providing a very interesting perspective in the medium and long term.
In particular, this endeavour would depend on a successful evaluation,
nonperturbative in $Z\alpha$, of all two-loop binding corrections
to the bound-electron $g$ factor.

As Fig.~\ref{fig5} shows, the determination of $\alpha$ based
on the bound-electron $g$ factor currently
appears much more promising for extremely light systems,
such as ${}^4{\rm He}^+$.
The measurement of $g$ factor, however, would definitely
have to be carried out with an accuracy better than
$10^{-11}$ in order to match the current accuracy
for $\alpha$. Alternatively
(see Fig.~\ref{fig4}), the planned $g$ factor measurement in
${}^{40}{\rm Ca}^{19+}$ could potentially lead to a
value of $\alpha$ that matches the accuracy of the free-electron value,
provided the two-loop uncertainty (higher-order binding corrections)
can be reduced and provided the accuracy of the atomic mass
determination can be enhanced beyond $10^{-10}$.
With current theory, the accuracy of the determination
of $\alpha$ from the ${}^{40}{\rm Ca}^{19+}$ measurements is limited
to an accuracy of about two orders of magnitude less than
the free-electron value.

A final word on the electron mass:
For a speculative alternative determination of $\alpha$ in a high-$Z$
experiment, the current accuracy of $m_{\rm e}$,
based on the carbon and oxygen
measurements~\cite{HaEtAl2000prl,BeEtAl2002prl,YeInSh2002,%
VeEtAl2004,PaJeYe2004,PaCzJeYe2005} is sufficient.
We recall the values
(evaluated using the most recent theory~\cite{PaCzJeYe2005})
\begin{align}
m_{\rm e}({}^{12} {\rm C}^{5+}) =& 0.000\,548\,579\,909\,32(29) \, {\rm u}\,,
\\[2ex]
m_{\rm e}({}^{16} {\rm O}^{7+}) =& 0.000\,548\,579\,909\,60(41) \, {\rm u}\,.
\end{align}
However, if an alternative determination of $\alpha$ via low-$Z$
measurements is pursued in earnest, then it becomes necessary
to improve the value of $m_{\rm e}$ beyond the $10^{-11}$ threshold.

%
%
\section{Conclusions}
\label{conclu}

In Sec.~\ref{atmmass}, we emphasize the importance of current high-precision
and upcoming ultra-high precision atomic mass measurements
for the determination of fundamental physical constants, in combination
with bound-electron $g$ factor measurements in hydrogenlike systems.
One of the celebrated achievements connected to $g$ factor measurements
lies in the improvement of the
accuracy of the electron mass by a factor of 4, as compared to the
previous value based on measurements involving protons and electrons
in Penning traps~\cite{FaDySc1995}.

The expansion of the bound-electron
$g$ factor in terms of the two most important parameters
in the non-recoil limit is discussed in Sec.~\ref{theory}.
These are the
loop expansion parameter $\alpha$ (the fine-structure constant) and
the Coulomb binding parameter $Z\alpha$, where $Z$ is the nuclear
charge number. Furthermore, in Sec.~\ref{asymp},
we analyze generalized Bethe logarithms, termed $\ln k_3$,
which are relevant for binding corrections to the $g$ factor, in the
limit of large principal quantum number (i.e., for highly excited
Rydberg states). The calculation of the result
$\ln k_3(8S) = 5.022\,275\,220$ (see Table~\ref{lnk3table}),
facilitates the analysis of the asymptotic limit. The discussion
is accompanied by a tentative proposal~\cite{QuPriv2005} 
for a double-resonance experiment,
to probe the bound-electron $g$ factor for highly excited Rydberg
states with a long lifetime.
In Sec.~\ref{nucl}, we discuss prospects for determinations of nuclear
properties, and of the fine-structure constant, based on
measurements in various ranges of the nuclear charge number.
An alternative
measurement of the fine-structure constant, of comparable
accuracy to the free-electron value, could be
accomplished via measurements at low $Z$, provided the
experimental accuracy of the $g$ factor can be pushed
beyond 1 part in $10^{11}$,
and provided the electron mass can be determined to sufficient
accuracy (see also Figs.~\ref{fig4} and~\ref{fig5}).
{\em A priori}, combined ultra-high precision measurements
in ${}^4 {\rm He}^+$ and ${}^{10} {\rm Be}^{3+}$
appear to provide for a viable approach,
provided the atomic mass measurements of ${}^4 {\rm He}$
and ${}^{10} {\rm Be}$
can reach comparable accuracy (now, the experimental
accuracy stands at $1.5$ parts in
$10^{11}$ for ${}^4 {\rm He}$, see Ref.~\cite{AuWaTh2003}).
The two measurements in He and Be could provide input
data for a coupled system of equations, to be solved for $\alpha$
and $m_{\rm e}$.

By contrast, considerable
further theoretical and experimental
progress (concerning, e.g., nuclear radii)
is required before any such endeavour could be
realized in the domain of high nuclear charges.
The prerequisites are outlined in Sec.~\ref{nucl}. We conclude that even in
the absence of this progress, prospective measurements at
higher $Z$ will yield a rather interesting verification of quantum
electrodynamics in the high-field domain.

%
%
\section*{Acknowledgements}

Valuable discussions with W.~Quint are gratefully acknowledged.
U.D.J. acknowledges support from the Deutsche Forschungsgemeinschaft
via the Heisenberg program. This work was supported by EU grant
No.~HPRI-CT-2001-50034 and by RFBR grant No.~04-02-17574. A.C. was
supported by the Science and Engineering Research Canada. V.A.Y.
acknowledges support by the foundation ``Dynasty''.

\appendix

%
%
\section{Nuclear Magnetic Susceptibility Correction}
\label{diamagpol}

It is well recognized that the nuclear polarizability 
can shift atomic energy levels or electronic 
$g$ factors \cite{NePlSo2002}.
Less well 
known is the influence of nuclear magnetic susceptibility $\beta_M$,
which can be significant for large Z-nuclei.
The effective interaction Hamiltonian which defines $\beta_M$ is
\begin{equation}  
\label{A1}
\delta H = \frac{\beta_M}{2}\,\vec B^2 \,,
\end{equation}
where $\vec B$ is the magnetic field at the nucleus. Here,
we estimate $\beta_M$ on the basis of simple assumptions.
In particular,
we assume that nucleus is a bound system of nonrelativistic 
nucleons. Therefore, the Hamiltonian in the external magnetic field is
\begin{equation}
H = \sum_{i=1}^Z 
\left(\frac{\vec\pi^2_i}{2\,m_{\rm p}} -
\mu_{\rm p}\,\vec\sigma_i\,\vec B\right)
+ \sum_{j=1}^N 
\left(\frac{\vec p^2_j}{2\,m_{\rm n}} -\mu_{\rm n}\,
\vec\sigma_j\,\vec B\right) 
+ H_I\,,
\end{equation}
where $H_I$ is the interaction Hamiltonian, which we assume to be 
$\vec B$-indepedent. The proton and neutron masses are denoted
by $m_{\rm p}$ and $m_{\rm n}$, respectively.
The term linear in $\vec B$ gives the nuclear magnetic moment.
The quadratic term from $\vec\pi^2 = (\vec p + e\,\vec A)^2$ \,,
\begin{equation}  
\delta H = \sum_{i=1}^Z \frac{e^2\,\vec A^2}{2\,m_{\rm p}} = 
\sum_{i=1}^Z 
\frac{e^2}{8\,m_{\rm p}}\,(\vec r_i\times\vec B)^2\,,
\end{equation} 
gives the magnetic suseptibility $\beta_M$ (we denote 
by $e$ the electron charge, $e = - |e|$). If we assume
that the quadrupole moment vanishes or is negligible, then
\begin{equation}
\label{A4}
\beta_M = \frac{e^2}{4\,m_{\rm p}}\,\frac{2}{3}\,
\sum_{i=1}^Z\,\langle r_i^2\rangle
= \frac{Z\,e^2}{6\,m_{\rm p}}\,\langle r^2\rangle \,,
\end{equation}
where  $\langle r^2\rangle$ is the mean square charge radius of the nucleus.

We now turn to the correction to the $g$ factor. The magnetic field in
Eq. (\ref{A1}) is a sum of an external magnetic field $\vec B_{\rm ext}$
and the field $\vec B_{\rm el}$ produced by the bound electrons. 
This leads to an effective additional interaction
of the electronic magnetic moment with the external magnetic field,
\begin{equation}
\delta H = \beta_M\,\vec B_{\rm ext}\cdot \vec B_{\rm el} \equiv 
-\vec\mu_{\rm ind}\cdot\vec B_{\rm el} 
\end{equation}
where $\vec \mu_{\rm ind}$ is the induced nuclear magnetic moment.
In the nonrelativistic limit, the magnetic dipole interaction of 
the nuclear magnetic moment $\mu_I$ with the electron in the $S$-state 
results in a hyperfine structure Hamiltonian
\begin{equation}
\delta H_{\rm hfs} = -\frac{e}{3\,m_{\rm e}}\,
\vec \mu_I\cdot \vec \sigma_e\,
\langle\delta^3(r_{\rm e})\rangle\,.
\end{equation}
It is now straightforward to write down 
the analogous Hamiltonian $\delta H$ 
which describes the interaction with the 
induced nuclear magnetic moment $\mu_{\rm ind}$,
\begin{equation}
\delta H = -\frac{e}{3\,m_{\rm e}}\,\vec \mu_{\rm ind}\cdot
\vec\sigma_{\rm e}\,\langle\delta^3(r_{\rm e})\rangle =
\frac{e}{3\,m_{\rm e}}\,\beta_M\,\vec B_{\rm ext}\cdot  \vec
\sigma_{\rm e}\,\langle\delta^3(r_{\rm e})\rangle
\equiv \delta g\,\left(\frac{-e}{2\,m_{\rm e}}\right)\,
\frac{\vec \sigma_{\rm e}}{2}\cdot\vec B_{\rm ext}
\end{equation}
where
\begin{equation}
\delta g = -\frac{4}{3}\,
\langle\delta^3(r_{\rm e})\rangle\,\beta_M
 = -\frac{4\,(Z\,\alpha)^3}{3\,\pi\,n^3}\,
m_{\rm e}^3\,\beta_M\,,
\end{equation}
and $\beta_M$ is defined in Eq. (\ref{A4}).
As an example, one may consider ${}^{40}{\rm Ca}$ with
$Z = 20$ and a radius of
$\sqrt{ \left< \vec{r}^2 \right> } = 3.4764(10) \, {\rm fm}$~\cite{An2004}.
Using values for the fundamental constants as given
in~\cite{MoTa2005}, one obtains an estimate for the nuclear
susceptibility of $\beta_M = 1.35\cdot 10^{-8}\,m_e^{-3}$ and
$\delta g_{\rm Ca} \approx -1.78 \cdot 10^{-11}$ for $n=1$,
which is much less than the uncertainty due to higher order two-loop
corrections but important for an accurate understanding 
of nuclear contributions to the bound-electron $g$ factor.

\end{document}